\documentclass[aps]{revtex4}
\usepackage{graphicx}
\usepackage{bm}
\usepackage{amsmath}
\usepackage[T1]{fontenc}
\usepackage{mathrsfs,ae,dsfont}

\begin{document}

\title{Scaling Transformation for Nonlocal Interactions}

\author{Hai-Jun Wang}

\address{Center for Theoretical Physics and School of Physics, Jilin
University, Changchun 130012, China}

\begin{abstract}
In the light of their relationships with renormalization, in this
paper we associate the scaling transformation with nonlocal
interactions. On one hand, the association leads us to interpret
the nonlocality with locally symmetric method. On the other hand,
we find that the nonlocal interaction between hadrons could be
test ground for scaling transformation if ascribing the running
effects in renormalization to scaling transformation. First we
derive directly from group theory the operator/coordinate
representation and unitary/spinor representation for scaling
transformation, then link them together by inquiring a
scaling-invariant interaction vertex mimicking the similar process
of Lorentz transformation applied to Dirac equation. The main
feature of this paper is that we discuss both the representations
in a sole physical frame. The representations correspond
respectively to the spatial freedom and the intrinsic freedom of
the same quantum system. And the latter is recognized to
contribute to spin angular momentum that in literature has never
been considered seriously. The nonlocal interaction Lagrangian
turns out to vary under scaling transformation, analogous to
running cases in renormalization. And the total Lagrangian becomes
scale invariant only under some extreme conditions. The
conservation law of this extreme Lagrangian is discussed and a
contribution named scalum appears to the spin angular momentum.
Finally a mechanism is designed to test the scaling effect on
nonlocal interaction.
\end{abstract}

\maketitle

\section{Introduction}

Nowadays, on account of the developments of string
theory~\cite{Ganor07, Min08}, Lattice QCD~\cite{Zol12} and the
necessity to describe nonperturbatively the intermediate strong
interaction between extended hadrons~\cite{Tim13}, the
construction of a consistent nonlocal theory is still called for
\cite{Eliezer89, Moffat89, Evens91, Kleppe92, Moffat90, Neves10,
Campo10, Br¨¹n04, Haque09, Dowker09, Barn11, Calca08, Hofman11,
Moffat11, Zhou08, Pimi10}. The pioneering study of nonlocal
interaction dates back to the 1930's ~\cite{Wata34} when quantum
field theory was in its infancy. And the phenomenology of nonlocal
interaction commenced with the primary attempts to describe the
interaction between extended particles (such as
hadrons~\cite{Yukawa50,Yennie50}), whilst to cope with the
divergence appearing in local quantum field theories (LQFT). The
development afterward purported mainly to give a consistently
convergent theory in order to underlie the named ''effective field
theory'', whereof some form factors were usually employed
\cite{Kristen52,Bloch50,Bloch52, Pauli53, Hayashi53, Lee57,
Dirac62, Kir66, Efimov67, Efimov72, Alebas73, Alebas73A, Marnel74,
Eliezer89, Moffat89, Evens91, Kleppe92, Moffat90}. Whereas in such
context one encounters the difficulty of unitarity and causality
in formulating the S matrix~\cite{Pauli53, Marnel74}, no matter
the Feldman-Yang~\cite{Yang50} method or conventionally canonical
quantization method ~\cite{Pauli53}. Some promising progresses on
this issue~\cite {Efimov72, Alebas73, Moffat89,Moffat90, Evens91,
Klep92, Klep93, Clay01, Calian00, BHA98,Wang11} in one way or
another showed their accordance with the renormalization
methods~\cite{Orland09,Orland11,Su99,Wang04}. In this paper we try
to use the scaling transformation (dilatation of space-time),
which is inspired by renormalization and assumed effective in
nonlocal description of hadron physics, to unveil part of running
effects in nonlocal interaction.

~~\\

More often than not, previous investigations on nonlocal
interaction tried to fit certain results to those of
renormalization~\cite{Efimov72, Alebas73, Moffat89,Moffat90}.
Reversely, in this paper we phenomenologically extract a scaling
transformation from renormalization for nonlocal interaction based
on their similar physical picture. With afterthought, the
achievements of finding that nonlocal interaction are linked with
renormalized interaction vertex may in one aspect owe to their
common characteristic of effectively using form factors. For
instance, in QED, the momentum-space vertex with renormalization
is \cite{Ian13,Arr11}
\begin{equation}
\Gamma ^\mu (p^{\prime },p)=\gamma ^\mu F_1(q^2)+\frac{i\sigma ^{\mu \nu
}q_\nu }{2m}F_2(q^2)\text{ .}  \label{renorm}
\end{equation}
where $q=p^{\prime }-p$, with $F_1(q^2)$ and $F_2(q^2)$ known as
Dirac and Pauli form factors respectively. Similarly, the nonlocal
interaction has its general form factors in coordinate-space
~\cite{Bloch50, Marnel74},

\begin{equation}
\mathscr{L}_I=-g\int \int d^4\xi d^4\eta F(\xi ,\eta )\,A_\mu (x)\bar \psi
(x+\eta )\Gamma ^\mu \psi (x+\xi )\text{ .}  \label{LagA}
\end{equation}
Here the vertex $\Gamma ^\mu $ could be the usual vector $\gamma ^\mu $,
tensor like $i\sigma ^{\mu \nu }q_\nu $, or other forms to be determined.
Its form in momentum-space then is
\[
\mathscr{L}_I=-gA_\mu (q^2)\bar \psi (p)F(p,p^{\prime })\Gamma ^\mu \psi
(p^{\prime }).
\]
where $q=p^{\prime }-p$, $\psi (p^{\prime })$ is the spinor in momentum
space and the expansion like $\psi (x+\xi )=\sum_{\mathbf{p}}\psi
(p)e^{ip(x+\xi )}$ has been implicit. The $F(p,p^{\prime })$ is defined as
the Fourier transform of $F(\xi ,\eta )$%
\begin{equation}
F(p,p^{\prime })=\int \int d^4\xi d^4\eta e^{ip\eta }F(\xi ,\eta
)\,e^{-ip^{\prime }\xi }\text{ .}  \label{Fourier}
\end{equation}

The renormalization group method (RGM) has its intrinsic
relationship with scaling transformation if viewing the
differentiating operator $\mu \frac{\text{d}}{\text{d}\mu }$ in
group equation as scaling operator. In RGM, for a function
$\Lambda$ that represents a vertex function, a wave function or a
propagator, its renormalized form and unrenormalized form are
linked as~\cite{Su99}
\[
\Lambda =Z_F\Lambda _R\text{ ,}
\]
whence the form factor $F(p,p^{\prime })$ may be (approximately)
viewed as just the collection of these $Z_F$s, which are obtained
by loop corrections. Differentiating the above equation with
respect to renormalization parameter $\mu $, and in view of that
unrenormalized $\Lambda $ is independent of $\mu $, one
immediately gets
\begin{equation}
\mu \frac{\text{d}\Lambda _R}{\text{d}\mu }+\gamma _F\,\Lambda _R=0\text{ ,}
\label{Scale A}
\end{equation}
where $\gamma _F$ is the anomalous scaling dimension defined by
\[
\gamma _F=\mu \frac{\text{d}}{\text{d}\mu }\ln Z_F\text{ .}
\]
In the next section one may note that the operator $\mu
\frac{\text{d}}{\text{d}\mu }$ is just the scaling operator in its
spatial representation, apart from a coefficient $i$. The eq.
(\ref{Scale A}) is a special form of renormalization group
equation, and the well known form is~ \cite{Cal70,Syman70}
\[
\lbrack M\frac \partial {\partial M}+\beta (\lambda )\frac \partial
{\partial \lambda }+n\gamma (\lambda )]G^{(n)}(\{x_i\};M,\lambda )=0
\]
which is for any Green's function of massless $\phi ^4$ theory.
Supposes the function $\Lambda _R$ has a dimension $\gamma _F$
with respect to a scale parameter $\mu $, then by such
transformation $\mu \rightarrow \lambda \mu $, the $\Lambda _R$
yields
\[
\Lambda _R(\mu ,\text{other parameters})=\lambda ^{\gamma
_F}\Lambda _R(\frac \mu \lambda ,\text{other parameters}) \text{
,}
\]
that is the essence of RGM. Besides the obvious application of
spatial scaling-transformation to the nonlocal form factor
$F(p,p^{\prime })$, in this paper we will focus on how it affects
spinors consistently while involving its unitary-representation.

~~\\

The scaling transformation, i.e. a freedom added to Poincare group
to form Wely group~\cite{Budi79}, belongs to a larger group called
4-dimension Conformal Group, which in mathematical side has been
investigated thoroughly from different aspects, and its
application to physics especially to quantum field once was also
widely considered. However the application is not so satisfactory
because hitherto no other perfect quantum system than photon field
\cite{Bate1910,Cunn1910} has been found so that the corresponding
Lagrangian is scaling invariant, i.e. demanding the mass of
involved particles to be null~\cite {Kast08, Yu13, Jose88,
Gross70, Dirac35,Lus75}. Furthermore, one inference of the scale
invariance is that according to Noether's theorem, if a Lagrangian
is invariant under scaling transformation, then the trace of the
energy-momentum tensor should be null~\cite{Kast08,Yu13}. These
two factors become obstacles to apply the scaling transformation
to most material fields. Other efforts were also experienced to
search for invariant fermion equation or scattering amplitude
~\cite{Dirac35, Gross70, Mack69}, and even to apply it to nonlocal
action ~\cite{Ryder74,Whe4549}. None of the results is pertinent
to known material fields. In this paper we investigate the
application by trying two new tentative methods. One is to
consider the unitary representation and the coordinate
representation of conformal group simultaneously. The other is to
apply the scale transformation to hadron physics since, the
hadrons have their own sizes with which the interactions between
them to some extent vary. Accordingly the test bed for conformal
transformation might be nonlocal interaction between hadrons.
However, here the scaling transformation works not for invariance,
but for running. The running effects of nonlocal interaction are
just like those in renormalization.

~~\\

In this paper we shall use the scaling feature of RGM, but we free
us from the detail calculation of renormalization. Since we are
looking for a transformation method to interpret the running
effect in nonlocal interaction, once we already got an effective
form of Lagrangian or Hamiltonian, we would just use tree-level
form to do calculations. The further loop calculation will double
count something, Born approximation is fine for most cases of
interest. The calculation resembles that used in deep inelastic
scattering, though we are involved just elastic processes. In
summary, in the whole paper we focus more on the properties of
scaling transformation/conformal group, and on how to apply them
to nonlocal fields.

~~\\

The rest of the paper is arranged as follows. Sect. II is
dedicated to introducing the two representations of scale
transformation, i.e. the coordinate/operator representation and
spinor/unitary representation. In Sect. III we establish the
physical relationship between the two representations, on
condition that a scale invariant vertex exists. Subsequently in
Sect. IV we discuss the conservation law for the derived
scaling-invariant vertex, and the possibility that it relates to
the nucleon's polarizations is posed. In Sect. V according to the
characteristics of applying the general vertex-form $\gamma ^\mu
(a\ +b\gamma ^5)$ to polarized scattering, a mechanism is proposed
to examine the predictions on nonlocality. Conclusions and
discussions are presented finally.

~\\

\section{The Spatial and Spinor Representations for Scaling Transformation
Based on Group Theory}

~~\\

It is well known that the scaling transformation belongs to a
larger Conformal Group~\cite {Budi79,Yu13}, therefore next we will
learn first the properties of 4-dimensional Conformal Group,
including its spatial/operator representation and unitary/spinor
representation, as well as commutations among their generators. At
the end of this section, we will understand the role of operator
$\mu \frac{\text{d}}{\text{d}\mu } $ in the conformal group. The
spatial representation is mainly referencing to that of Ref.~\cite
{Budi79A, Budi79} and the unitary representation is derived by
applying Cartan method ~\cite{Cartan37} to $SO(6)-SU(4)$
transform. The unitary representation is the focus of this
section, and of this paper as well.

Mostly the scaling transformation in 4-dimension is discussed as a subset of
conformal group, and in previous literature its applications are seldom
considered independently~\cite{Yu13}. Here we start with the null vector
space (Euclidean space),
\begin{equation}
\eta_1^2+\eta_2^2+\eta_3^2+\eta_4^2+\eta_5^2+\eta_6^2=0\text{ .}
\label{LengthinReal}
\end{equation}
reserving which gives the popular definition of conformal
group~\cite {Cartan37}. A special expression of the differential
forms in 4-dimension spatial representation can be derived
directly from the above equation. In derivation we need to apply
the following variables \cite{Budi79}
\begin{equation}
x_\mu =\frac{\eta _\mu }K\text{ , where }K=\eta _5+i\,\eta _6\text{ , where }%
\mu =1,2,3,4  \label{projection}
\end{equation}
together with the differential form
\begin{equation}
\frac \partial {\partial \eta _a}=\frac 1K\{[\delta _{a\mu }-(\delta
_{a5}+i\delta _{a6})x_\mu ]\frac \partial {\partial x_\mu }+(\delta
_{a5}+i\delta _{a6})K\frac \partial {\partial K}\}\text{ , where }%
a=1,2,\cdots ,6  \label{aa}
\end{equation}
to the definition of 6-dimensional angular-momentum
\begin{equation}
M_{ab}=i(\eta _a\frac \partial {\partial \eta _b}-\eta _b\frac \partial
{\partial \eta _a})\text{, where }a, b=1,2,\cdots ,6\text{ .}  \label{bb}
\end{equation}
Then one gets the following generators for conformal group \cite{Budi79} [of
which in eq. (56)]
\begin{eqnarray}
D &=&iM_{56}=-(\eta _5\frac \partial {\partial \eta _6}-\eta _6\frac
\partial {\partial \eta _5})=i(x_\mu \frac \partial {\partial x_\mu
}-K\frac \partial {\partial K})\text{,}  \nonumber \\
P_\mu &=&M_{5\mu }+iM_{6\mu }=i\frac \partial {\partial x_\mu }\text{ , }%
K_\mu =M_{5\mu }-iM_{6\mu }=i\{-x^2\frac \partial {\partial x_\mu
}+2x_\mu x_\nu \frac \partial {\partial x_\nu }-2K\,x_\mu \frac
\partial {\partial K}\}\text{,}  \label{cc}
\end{eqnarray}
The projected form (making $K$ as constant boundary of Minkowski
space\cite {Ma99}) with Minkowski convention then is
\begin{eqnarray}
D &=&i\,x_\mu \frac \partial {\partial x_\mu }\text{, }M_{\mu \nu }=i(x_\mu
\frac \partial {\partial x^\nu }-x_\nu \frac \partial {\partial x^\mu })%
\text{,}  \nonumber \\
P_\mu &=&i\frac \partial {\partial x^\mu }\text{ , }K_\mu =-i(x^2\frac
\partial {\partial x^\mu }-2x_\mu x^\nu \frac \partial {\partial x^\nu })%
\text{,}  \label{ccc}
\end{eqnarray}
where $M_{\mu \nu }$ represent the components of conventional
angular momentum in 4-dimension. The corresponding commutation
relation can be obtained by direct computation,
\begin{eqnarray}
\lbrack M_{\mu \nu },M_{\rho \sigma }] &=&i(g_{\nu \rho }M_{\mu \sigma
}+g_{\mu \sigma }M_{\nu \rho }-g_{\mu \rho }M_{\nu \sigma }-g_{\nu \sigma
}M_{\mu \rho }),  \nonumber \\
\lbrack M_{\mu \nu },P_\rho ] &=&i(g_{\nu \rho }P_\mu -g_{\mu \rho }P_\nu ),
\nonumber \\
\lbrack D,P_\mu ] &=&-iP_\mu \text{, }[D,K_\mu ]=iK_\mu ,  \nonumber \\
\lbrack D,M_{\mu \nu }] &=&0  \nonumber \\
&&\ \ \cdots \cdots \cdots \cdots  \label{dd}
\end{eqnarray}

Before using Cartan method to achieve the unitary representation
of Conformal Group, let's review first the steps of Cartan method
with $SO(3)-SU(2)$ mapping as an
example \cite{Cartan37}[of which in pp. 41-48]. To keep the invariance of $%
x_1^2+x_2^2+x_3^2=0$, one defines the matrix
\begin{equation}
X=\left(
\begin{array}{cc}
x_3 & x_1-i\,x_2 \\
x_1+i\,x_2 & -x_3
\end{array}
\right) \text{ .}  \label{cc}
\end{equation}
The trace Tr$(X^{\dagger }X)$ is $x_1^2+x_2^2+x_3^2$. With $U$ as
an element of $SU(2)$ group, we define
\begin{equation}
X^{\prime }=U^{-1}XU\text{ ,}  \label{ee}
\end{equation}
immediately we have
\begin{equation}
Tr(X^{\prime \dagger }X^{\prime })=Tr(X^{\dagger }X)\text{ ,}
\label{xx}
\end{equation}
thus $SU(2)$ group keeps the trace invariant, and by this way the
group also keeps the metric $x_1^2+x_2^2+x_3^2$. With the
knowledge that the $SO(3)$ group directly reserves the metric
$x_1^2+x_2^2+x_3^2$, we conclude that Cartan matrix $X$ acts as a
mapping between $SO(3)$ and $SU(2)$. By the Cartan Matrix $X$, one
can define spinor $\left(
\begin{array}{c}
\xi _0 \\
\xi _1
\end{array}
\right) $ by
\begin{equation}
X\,\left(
\begin{array}{c}
\xi _0 \\
\xi _1
\end{array}
\right) =0\text{ ,}  \label{ks}
\end{equation}
with the solution $\xi _0=\pm \sqrt{\frac{x_1-i\,x_2}2}$ and $\xi
_1=\pm \sqrt{\frac{-x_1-i\,x_2}2}$, and the reverse yields
\begin{eqnarray}
x_1 &=&\xi _0^2-\xi _1^2  \nonumber \\
x_2 &=&i\,(\xi _0^2+\xi _1^2)  \nonumber \\
x_3 &=&-2\xi _0\xi _1\text{ ,}  \label{spinor}
\end{eqnarray}
which automatically satisfies $x_1^2+x_2^2+x_3^2=0$ from which we can define
the spinor reversely.

From the above Cartan matrix $X$ we can extract the Pauli matrices $\sigma
_1 $, $\sigma _2$, $\sigma _3$ separately from the coefficients of $x_1$, $%
x_2$, $x_3$. Meanwhile Pauli matrices $\sigma _1$, $\sigma _2$,
$\sigma _3$ act as the generators of $SU(2)$ group mentioned
above. Furthermore it is easy to test that $SU(2)$ group reserves
the metric
\begin{equation}
\mid \xi _0\mid ^2+\mid \xi _1\mid ^2=\Xi ^{\dagger }\Xi \text{ .}
\label{Dspinor}
\end{equation}
And coincidentally the $n-$vectors form (defined in eq.
(\ref{k-vector})) based on Pauli matrices don't generate new
matrices, neither the multiplications nor the commutations among
them, they themselves are closed. Now in what follows we would
find the corresponding Cartan matrix from $ SO(6)$ to
$SU(4)$/$SU(2,2)$, namely the spinor representation for
4-dimension Conformal group.

To achieve its unitary/spinor representation in 4-dimension, mimicking the
relationship between the metric $x_1^2+x_2^2+x_3^2$ and that in Eq. (\ref
{Dspinor}), we shall associate the metric in Eq. (\ref{LengthinReal}) with
the invariant quadratic form
\begin{equation}
\mid z_1\mid ^2+\mid z_2\mid ^2+\mid z_3\mid ^2+\mid z_4\mid ^2=Z^{\dagger }Z%
\text{ ,}  \label{UniLength}
\end{equation}
by the following matrix~~\cite{Este64},
\begin{equation}
A=\left(
\begin{array}{cccc}
0 & x_1+i\,x_2 & x_3+i\,x_4 & x_5+i\,x_6 \\
-(x_1+i\,x_2) & 0 & x_5-i\,x_6 & -x_3+i\,x_4 \\
-(x_3+i\,x_4) & -x_5+i\,x_6 & 0 & x_1-i\,x_2 \\
-(x_5+i\,x_6) & x_3-i\,x_4 & -x_1+i\,x_2 & 0
\end{array}
\right) \text{ .}  \label{MatrixCorres}
\end{equation}
Count the degrees of freedom of the groups that conserv separately
Eq. (\ref {LengthinReal}) and Eq. (\ref{UniLength}), one finds
they are both 15. Next we only need to extract the coefficients
before $x_i$'s to get the unitary matrices as generators of
$SU(4)$, just like the method used in three dimension example~Eqs.
(\ref{dd}-\ref{spinor}). If we want to get the
generators of $SU(2,2)$ we need only to change the signs before $x_1$ and $%
x_2 $ and those ahead of corresponding matrices, which would change the eqs.
(\ref{LengthinReal}) and (\ref{UniLength}) to
\begin{equation}
-x_1^2-x_2^2+x_3^2+x_4^2+x_5^2+x_6^2=0\ \text{ .}  \label{LengthinRealB}
\end{equation}
and
\begin{equation}
-\mid z_1\mid ^2-\mid z_2\mid ^2+\mid z_3\mid ^2+\mid z_4\mid ^2=Z^{\dagger
}Z\text{ .}  \label{UniLengthB}
\end{equation}
the latter falls into Dirac spinor like
\[
\tilde \psi =(z_1,z_2,z_3,z_4)\text{ .}
\]
It can be examined that the matrix $A$ in Eq. (\ref{MatrixCorres}) meets the
invariant expression
\begin{equation}
Tr(A^{\dagger }A)=4(x_1^2+x_2^2+x_3^2+x_4^2+x_5^2+x_6^2)  \label{trace}
\end{equation}
just like the above 3-dimension example, while the $SU(4)$ group
keeps the above trace
$x_1^2+x_2^2+x_3^2+x_4^2+x_5^2+x_6^2=constant$, it simultaneously
reserves the metric Eq. (\ref{UniLength}). The above method of
linking real metric to a matrix is closely analogous to the Cartan
method of constructing a spinor representation in any real space.
Actually, the true spinor space for 4-d conformal group following
Cartan method should be of 8-dimension instead of 4-dimension
\cite{Cartan37}[of which in pp. 88-89]. In what follows we would
take over the process of deriving all of the $n$-vectors along the
Cartan method~\cite{Cartan37}[of which in pp.81-83], though we
work in 4-dimension rather than 8-dimension. First we extract the
matrices before $x_i$'s in Eq. (\ref{MatrixCorres}) ,
$i.e.1-$vectors,
\begin{eqnarray}
B_1 &=&\left(
\begin{array}{cc}
i\,\sigma _2 & 0 \\
0 & i\,\sigma _2
\end{array}
\right) \text{,}\;B_2=\left(
\begin{array}{cc}
-\,\sigma _2 & 0 \\
0 & \,\sigma _2
\end{array}
\right) \text{ ,}  \nonumber  \label{1vector} \\
B_3 &=&\left(
\begin{array}{cc}
0 & \,\sigma _3 \\
-\,\,\sigma _3 & 0
\end{array}
\right) \text{,}\;B_4=\left(
\begin{array}{cc}
0 & i\,I \\
-i\,I & \,0
\end{array}
\right) \text{ ,}  \nonumber \\
B_5 &=&\left(
\begin{array}{cc}
0 & \,\sigma _1 \\
-\,\,\sigma _1 & 0
\end{array}
\right) \text{,}\;B_6=\left(
\begin{array}{cc}
0 & \,-\sigma _2 \\
-\,\,\sigma _2 & 0
\end{array}
\right) \text{ .}  \label{1vector}
\end{eqnarray}
where $\sigma _i$'s are Pauli matrices. The definition of $k$-vector is
\begin{equation}
B_{k-vector}=\sum_P(-1)^PB_{n_1}B_{n_2}\cdots B_{n_k}\text{ ,}
\label{k-vector}
\end{equation}
where $P$ denotes different permutations. Apply the above fomula to
2-vector, and use the corresponding subscripts to denote the $1$-vectors
involved, then
\[
B_{12}=B_1B_2-B_2B_1=\left(
\begin{array}{cc}
i\,\sigma _2 & 0 \\
0 & i\,\sigma _2
\end{array}
\right) \left(
\begin{array}{cc}
-\,\sigma _2 & 0 \\
0 & \,\sigma _2
\end{array}
\right) -\left(
\begin{array}{cc}
-\,\sigma _2 & 0 \\
0 & \,\sigma _2
\end{array}
\right) \left(
\begin{array}{cc}
i\,\sigma _2 & 0 \\
0 & i\,\sigma _2
\end{array}
\right) =0\text{ .}
\]
Similarly, let's exhaust all possibilities, then obtain other nontrivial
2-vectors
\begin{eqnarray}
B_{13} &=&2\left(
\begin{array}{cc}
0 & -\,\sigma _1 \\
\,\,\sigma _1 & 0
\end{array}
\right) \text{ , }B_{15}=2\left(
\begin{array}{cc}
0 & \,\sigma _3 \\
\,-\,\sigma _3 & 0
\end{array}
\right) \text{, }  \nonumber  \label{sigma1} \\
B_{35} &=&2i\left(
\begin{array}{cc}
\,\sigma _2 & 0 \\
0 & \,\sigma _2
\end{array}
\right) \text{, }B_{36}=2i\left(
\begin{array}{cc}
\,\sigma _1 & 0 \\
0 & \,-\sigma _1
\end{array}
\right) \text{, }  \nonumber \\
B_{46} &=&-2i\left(
\begin{array}{cc}
\,\sigma _2 & 0 \\
0 & \,-\sigma _2
\end{array}
\right) \text{, }B_{24}=2i\left(
\begin{array}{cc}
0 & \,\sigma _2 \\
\,\,\sigma _2 & 0
\end{array}
\right) \text{ ,}  \nonumber  \label{sigma1} \\
\text{ }B_{23} &=&-2i\left(
\begin{array}{cc}
0 & \,\sigma _1 \\
\,\,\sigma _1 & 0
\end{array}
\right) \text{. }  \label{sigma1}
\end{eqnarray}
We note that the new ones which are independent of $B_i$'s are just $B_{23}$
and $B_{36}$. The same line can be followed to carry out the 3-vectors.
Ignoring the repeating ones, we find the new 3-vectors independent of both
1-vectors and 2-vectors are
\begin{eqnarray}
B_{123} &\sim &\left(
\begin{array}{cc}
0 & \,\sigma _3 \\
\,\,\sigma _3 & 0
\end{array}
\right) \text{, }B_{134}\sim \left(
\begin{array}{cc}
\,\sigma _1 & 0 \\
0 & \,\sigma _1
\end{array}
\right) \text{, }  \nonumber \\
B_{145} &\sim &\left(
\begin{array}{cc}
\,\sigma _3 & 0 \\
0 & \,\sigma _3
\end{array}
\right) \text{, }B_{245}\sim \left(
\begin{array}{cc}
\,\sigma _3 & 0 \\
0 & -\,\sigma _3
\end{array}
\right) \text{, }  \nonumber \\
B_{345} &\sim &\left(
\begin{array}{cc}
0 & \,\sigma _2 \\
\,-\,\sigma _2 & 0
\end{array}
\right) \text{, }B_{146}\sim \left(
\begin{array}{cc}
I & \,0 \\
0 & -I
\end{array}
\right) \text{, }  \nonumber \\
B_{124} &\sim &\left(
\begin{array}{cc}
0 & \,I \\
\,\,I & 0
\end{array}
\right) \text{ .}  \label{sigma2}
\end{eqnarray}
Computing the 4-vectors and the higher ones would not give new independent
matrices. Finally, we can rearrange all above k-vector-produced matrices as
follows ~\cite{Este64},
\begin{eqnarray}
U_i &=&\frac 12\left(
\begin{array}{cc}
\,\sigma _i & 0 \\
0 & \,\sigma _i
\end{array}
\right)  \nonumber \\
V_\mu &=&-\frac 12\left(
\begin{array}{cc}
\,\sigma _\mu & 0 \\
0 & \,-\sigma _\mu
\end{array}
\right)  \nonumber \\
W_\mu &=&\frac i2\left(
\begin{array}{cc}
\,0 & \sigma _\mu \\
\sigma _\mu & \,0
\end{array}
\right)  \nonumber \\
Y_\mu &=&\frac 12\left(
\begin{array}{cc}
\,0 & \sigma _\mu \\
-\sigma _\mu & \,0
\end{array}
\right)  \label{sigma3}
\end{eqnarray}
where $\sigma _i$, $i=1,2,3$ are normal Pauli matrices and $\sigma _0=\left(
\begin{array}{cc}
1 & 0 \\
0 & \,1
\end{array}
\right) $. The convention can be changed from Minkowski to
Euclidean spaces while instead requiring $\sigma _\mu ^2=-1$, i.e.
making $\sigma _0=i$ and replacing definition of $\sigma _i$ by
those in \cite{Este64}.

The route of inquiring the concrete matrices following Cartan
method as above could be a shortcut that rarely mentioned in
literature. It is can be checked that the commutations among
$U_i$, $V_\mu $, $W_\mu $, $Y_\mu $ are just those for conformal
group \cite{Budi79, Budi79A}, accordingly the mapping from these
matrices to corresponding differential-forms turns out to be
\begin{eqnarray}
U_i &\leftrightarrow &\gamma _i\gamma _j\longrightarrow i(x_j\frac \partial
{\partial x^k}-x_k\frac \partial {\partial x^j})\longrightarrow M_{jk}
\nonumber \\
W_i &\leftrightarrow &\gamma _0\gamma _i\longrightarrow i(x_i\frac \partial
{\partial x^0}-x_0\frac \partial {\partial x^i})\longrightarrow M_{0k}
\nonumber \\
W_0 &\leftrightarrow &\gamma _5\longrightarrow i\,x_\mu \frac \partial
{\partial x_\mu }\longrightarrow D  \nonumber \\
V_\mu +Y_\mu &\leftrightarrow &\gamma _\mu (1\pm \gamma _5)\longrightarrow
i\frac \partial {\partial x^\mu }\longrightarrow P_\mu  \nonumber \\
V_\mu -Y_\mu &\leftrightarrow &\gamma _\mu (1\mp \gamma _5)\longrightarrow
-i(\frac 12x_\nu x^\nu \frac \partial {\partial x_\mu }-x_\mu x_\nu \frac
\partial {\partial x_\nu })\text{ }\longrightarrow K_\mu \text{.}
\label{mapping}
\end{eqnarray}
We use $\longrightarrow $ to represent the accurate mappings and $%
\leftrightarrow $ the equivalence, and the commutations have been
examined by computer. Now we recognize that the role of operator $\mu \frac{\text{d}}{%
\text{d}\mu }$ (or $x_\mu \frac \partial {\partial x_\mu }$) in
the conformal group is equivalent to the scaling operator $D$,
with its unitary form $\gamma _5$.

\section{The physical relationship between the two representations of
scaling transformation}

Enlightened by Lorentz transformation, in this section we try to
link physically the spatial form of scaling transformation with
its spinor/unitary form, the former representing the realistic
expansions and contractions of space-time (dilatation and
shrinkage means the same), the latter representing the intrinsic
freedom very like spin angular momentum. Considering both
representations in a sole frame is the main feature of this paper.

~~\\

As for a nonlocal interaction $F(q^2)A^\nu (q^2)\bar \psi
(p)\gamma _\nu \psi (p^{\prime })$, besides knowing that the form
factor $F(q^2)$ runs with scaling parameter as described Eq.
(\ref{Scale A}), we are also curious about how a nonlocal
interaction vertex $\gamma _\mu $ varies with scale. Before
drawing any conclusion, let's first find the invariant vertex
$\Gamma _\mu $ under the scaling transformation by mimicking the
method of utilizing Lorentz transformation to Dirac equation. In
this way we link its spatial form with its spinor form. As for
Lorentz transformation, the transformation matrix $\left( \Lambda
_{\;\,\mu }^\nu \right) $ for $j^\mu (y)=\bar \psi
(y)\gamma ^\mu \psi (y)$ corresponds to a complex transformation $S$ for $%
\psi (y)$ so that the effect of the transformed result $\bar \psi
(y)S^{-1}\gamma ^\mu S\psi (y)$ is equivalent to $\bar \psi
(y)\Lambda _{\;\,\nu }^\mu \gamma ^\nu \psi (y)$. Referencing the
case of Lorentz transformation, our goal in this section is to
search for the corresponding vertex-form $\Gamma ^\mu $ so that it
links with transformation $S^{\prime }$ by $S^{\prime -1}\Gamma
^\mu S^{\prime }=\Lambda _{\;\,\nu }^{\prime \mu }\Gamma ^\nu $,
where $S^{\prime }=e^{\frac u2\gamma _5}$, $\gamma _5$ is the
spinor representation of the scaling operator $D$, and $\Lambda
_{\;\,\nu }^{\prime \mu }$ represent tensor's components of
scaling transformation.

Usually we perform the spatial Lorentz transformation on the
vectors $A_\mu $ and $\gamma ^\mu $. Obviously this combination
brings about invariant formalism like $A^\nu (q^2)\bar \psi
(p)\gamma _\nu \psi (p^{\prime })$. We follow the convention that
the same set of $\{\gamma ^\mu \}$ is used in different coordinate
systems, which naturally yields an equivalence transformation $S$
satisfying~\cite{Mandl10,Feynman62}
\begin{equation}
S^{-1}\gamma ^\mu S=\Lambda _{\;\nu }^\mu \gamma ^\nu \,=\gamma ^{\prime \mu
},  \label{EquTran1}
\end{equation}
where $\Lambda _{\;\nu }^\mu $ stand for the tensors' components
of the Lorentz transformation. Substituting the Eq.
(\ref{EquTran1}) into $A_\mu (x)\bar \psi (x)\gamma ^\mu \psi (x)$
yields
\begin{equation}
A_\mu ^{\prime }(y)\bar \psi ^{\prime }(y)S^{-1}\gamma ^\mu S\psi ^{\prime
}(y)=A_\mu ^{\prime }(y)\bar \psi (y)\gamma ^{\prime \mu }\psi (y)\text{%
\thinspace .}  \label{EquTran2}
\end{equation}

While looking for $\Gamma ^\mu $ we would follow the same
convention as that in the above paragraph, i.e., in different
coordinate system we use the same set of $\{\Gamma ^\mu \}$. Then
analogously, we use the form of the above formula Eq.
(\ref{EquTran1}) for scaling transformation as
\begin{equation}
S^{\prime -1}\Gamma ^\mu S^{\prime }=\Lambda _{\;\,\nu }^{\prime \mu }\Gamma
^\nu \text{\thinspace ,}  \label{SpinorCoodA}
\end{equation}
where formally we have used $\Lambda _{\;\,\nu }^{\prime \mu }$ to
represent the scaling transformation to every coordinate component
~\cite{Dirac35,
Gross70},~\cite{ Mack69}[of which eq.(2)] instead of using the usual form $%
e^{-\alpha }$\cite{Budi79A}. Slightly different from the operator $\mu \frac{\text{d}%
}{\text{d}\mu }$ appearing in renormalization group equation, here the
operator $D$ has the usual form $D=i\,x^\nu \partial _\nu $ , being a hermit
one. With the relation $e^{-i\,\alpha \,D}p_\mu e^{i\,\alpha \,D}=e^{-\alpha
}p_\mu $, i.e. $[D,\,p_\mu ]=-i\,p_\mu $\cite{Budi79A}, we have
\begin{equation}
(\Gamma ^\mu p_\mu )_{scaling\text{ }transform}^{\prime }=S^{\prime
-1}\Gamma ^\mu S^{\prime }\Lambda _{\;\,\mu }^{\prime \nu }\,p_\nu \text{%
\thinspace }=S^{\prime -1}\Gamma ^\mu S^{\prime }\;e^{-i\,\alpha \,D}p_\mu
e^{i\,\alpha \,D}\text{.}  \label{SpinorCoodB}
\end{equation}
Now let's submit $S^{\prime }=e^{\frac u2\gamma _5}$ obtained from
the last section, where $u$ is the infinitesimal parameter.
Formally we get
\begin{eqnarray}
S^{\prime -1}\Gamma ^\mu S^{\prime }\Lambda _{\;\,\mu }^{\prime \nu }\,p_\nu
&=&\,e^{-\frac u2\gamma _5}\Gamma ^\mu e^{\frac u2\gamma _5}(p_\mu )_{scaling%
\text{ }transform}^{\prime }  \nonumber  \label{SpinorCoodC} \\
&=&\,e^{-\frac u2\gamma _5}\Gamma ^\mu e^{\frac u2\gamma _5}e^{-i\,\alpha
\,D}p_\mu e^{i\,\alpha \,D}  \nonumber \\
&=&e^{-\frac u2\gamma _5}\Gamma ^\mu e^{\frac u2\gamma _5}e^{-\alpha }p_\mu
\text{ }  \nonumber  \label{SpinorCoodC} \\
&\doteq &\,e^{-\frac u2\gamma _5}\Gamma ^\mu e^{\frac u2\gamma _5}p_\mu
(1-\alpha )\text{ .}  \label{SpinorCoodC}
\end{eqnarray}
From the experience of calculating $\gamma -$matrix and the following
relations
\begin{equation}
e^{-\frac u2\gamma _5}\gamma ^\mu e^{\frac u2\gamma _5}\simeq (1-\frac
u2\gamma _5)\gamma ^\mu (1+\frac u2\gamma _5)\simeq \gamma ^\mu +u\,\gamma
^\mu \gamma _5\,,  \label{ConcreteA}
\end{equation}

\begin{equation}  \label{ConcreteB}
e^{-\frac u2\gamma _5}\gamma ^\mu \gamma _5e^{\frac u2\gamma _5}\simeq
(1-\frac u2\gamma _5)\gamma ^\mu \gamma _5(1+\frac u2\gamma _5)\simeq \gamma
^\mu \gamma _5+u\,\gamma ^\mu \,,
\end{equation}

\begin{equation}
e^{-\frac u2\gamma _5}\gamma ^\mu (1\pm \gamma _5)e^{\frac u2\gamma
_5}\simeq (1-\frac u2\gamma _5)\gamma ^\mu (1\pm \gamma _5)(1+\frac u2\gamma
_5)\simeq (1\pm u\,)\gamma ^\mu (1\pm \gamma _5)\,,  \label{ConcreteC}
\end{equation}
we find out a possible form of $\Gamma ^\mu $%
\begin{equation}
\Gamma ^\mu =\gamma ^\mu (1\pm \gamma _5)\text{ while }\alpha \sim u\text{ .}
\label{Important}
\end{equation}
The coefficients $(1\pm u\,)$ of Eq. (\ref{ConcreteC}) can be
contracted now to be $1$ with coefficients $(1\mp u\,)$ that come
from the transformation of $ p_\mu $. And we note that the
infinitesimal parameters $u$ and $\alpha $ are not independent. By
this way we set up the relationship between the operator $D$ and
its unitary counterpart $S^{\prime }=e^{\frac u2\gamma _5}$
directly.

~~\\
The key element of linking the two operators $D$ and $S^{\prime
}=e^{\frac u2\gamma _5}$ is the scale-invariant vertex $\Gamma
^\mu$. One notes that $S^{\prime }=e^{\frac u2\gamma _5}$ is
responsible for acting on Dirac spinor as expected, or
equivalently on the vertex $\Gamma ^\mu$. And the operator $D$ is
responsible for acting on the real vector coupling to $\Gamma
^\mu$. Thus the scaling invariance holds true for interaction
vertex $\Gamma ^\mu \,A_\mu $, as well as for $\Gamma ^\mu \,p_\mu
$. The resultant vertex $\Gamma ^\mu=\gamma ^\mu (1\pm \gamma _5)$
is different from that of Ref. \cite{Mack69} due to the choice of
$\gamma _5$, since we have followed the convention of Quantum
Field Theory. All in all, we have extended transformation,
interaction vertex and spinor space simultaneously, which is
reasonable from the viewpoint of entirety.
~~\\

Now we are interested in what if we perform the scaling transformation $%
S^{\prime }$ succeedingly $N$ times upon the vector vertex-form
$\gamma ^\mu $. How the vector vertex form $\gamma ^\mu $ varies
with scaling is the starting point as well as the end of this
research. Different from Eqs. (\ref {ConcreteA}, \ref{ConcreteB},
\ref{ConcreteC}), now we employ the following formulism without
approximation
\begin{equation}
(e^{-\frac u2\gamma _5})^N\gamma ^\mu (e^{\frac u2\gamma _5})^N=\gamma ^\mu
\cosh Nu+\gamma ^\mu \gamma _5\sinh Nu\,,  \label{Correct}
\end{equation}
from which one notes that the vector vertex arrives at its limits
$\gamma ^\mu (1\pm \gamma _5)$ only if $\frac{\cosh Nu}{\sinh
Nu}\rightarrow \pm 1$, i.e. $Nu$ $\rightarrow \pm \infty $. $Nu$
$\rightarrow \pm \infty $ means one carrying out enough steps of
inflating or shrinking transformation. We call such states that
involve interaction vertices $\gamma ^\mu (1\pm
\gamma _5)$ as extreme states, which evolve from the interaction vertex $%
\gamma ^\mu $ after the scale constantly changing. And the
variation of coupling constant is assumed to be absorbed into
$F(q^2)$. It turns out that such scaling transformation doesn't
conserve the vector-dominant interaction, or alternatively, the
transformation tends to transform the relating spinor from a
normal one to a chiral one.

~~\\

Apart from these two extremes, the true vertex-form for nonlocal
interaction would mostly be of mixture form like $a\gamma ^\mu
+b\,\gamma ^\mu \gamma _5$ after carrying finite steps of scaling
transformation. The physics picture could be understood as follows
[Fig.1]. Initially, the pure vector-form $ \gamma ^\mu$ plays a
rough role in describing the interaction between a point particle
and an extended particle. As for the extended particle, while the
interaction is very weak i.e. the interaction energy is very low,
obviously it looks approximately like a point particle, i.e. not a
physical particle. So $\gamma ^\mu$ marks initially the rough
interaction between two point-particles. Now let's zoom in, i.e.
improving the energy (momentum) of interaction, then we can
imagine that the extended particle becomes gradually sizable in
contrast to original point-like. "Zooming in" is equivalent to, as
we propose here, many steps of scaling transformation. After
finite steps of transformation, the initial vertex $\gamma ^\mu$
would somehow evolve to a mixture form $a\gamma ^\mu +b\,\gamma
^\mu \gamma _5$, with which one can use local vertex-form and form
factor to interpret nonlocal interaction on certain energy scale.
And the additional coefficient $a$ is assumed to be part of the
form factor $F(q^2)$ of vector interaction, thus equivalent to the
running of coupling constant. This picture coincides with that of
renormalization. The conclusion of the above paragraphes also
tells that while the initial interaction between points being
$\gamma ^\mu \pm\,\gamma ^\mu \gamma _5$, then while we zooming
in, the interaction between the point and the true extended
particle would not change.  The extreme form $\gamma ^\mu
\pm\,\gamma ^\mu \gamma _5$ between points are the particular
cases that seldom occur. The weak interaction between neutrinos
and leptons belongs to such category.

~~\\

\section{The conservation law for the scale-invariant interaction $A_\mu
^{\prime }(x,y)\bar \psi (y)\gamma ^\mu (1\pm \gamma _5)\psi (y)$}

The necessity of studying the conservation law for the extreme
vertices is that such vertices might exist for a very short moment
in some scattering processes. According to Eq. (\ref{Correct}), to
repeatedly perform the transformation succeedingly until $\mid
Nu\mid $ becomes very large, the incident particle would approach
to a very high energy (or a very low energy) and its wave shrinks
(inflates) to a very small scale (a very large scale). At such
very high (low) energy scale, it is hard for the particles to
shrink (inflate) more, and its interaction vertex gets to the form
$A_\mu (x,y)\bar \psi (y)\gamma ^\mu (1\pm \gamma _5)\psi (y)$.
This interaction vertex may appear to systems of two hadrons
colliding at a very high (low) energy and exist just for a very
short instant of time, though not matching any true fundamental
interactions. $\gamma ^\mu (1\pm \gamma _5)$ make their sense
relative to their original form $\gamma ^\mu $---they have evolved
from the vertex-form $\gamma ^\mu $. $\gamma ^\mu (1\pm \gamma
_5)$ describes the nonlocal interaction between a point particle
and an extended particle, whereas $\gamma ^\mu $ underlies the
local interaction between the point particle and a point in
extended hadron. The deviation of the vertex-form $\gamma ^\mu
(1\pm \gamma _5)$ from vector $\gamma ^\mu $ suggests that a new
conservative current may appear~\cite{LurBook}. In what follows we
will study what might be the conservation law for the angular
momentum of such a nonlocal system at its extreme state, as well
as the impact of such conservation law on scattering processes
between extended particles.

From a classical point of view, while a ''soft'' body (with definite mass $m$%
) rotating, its shrinkage or inflation (like zooming in or out)
would not alter its total orbital angular momentum. However for a
quantum particle, its shrinkage or inflation occurs only when it
absorbs or releases a certain amount of energy. Such kind of
energy exchange of course breaks the angular-momentum conservation
by intuition. But this intuition is right only partially, since in
what follows we recognize that only spin part is varied, and the
spatial angular momentum is not varied due to the commutation
$[D,M_{\mu \nu }]=0$ \cite{Mack69}. The case is similar to that
when we extend three-dimensional rotation to four-dimensional
rotation, whereby we find the 3-dimensional orbital angular
momentum is not a conservative quantity any longer, unless we
further include the spin angular momentum. Now with scaling
transformation, we find the sum of orbital angular momentum and
the spin is not conserved any longer, so we have to include the
named ''scalum'' to find a conserved quantity. The ''scalum''
should be manifested by the transformation of spinors. In such
sequence we call the scaling transformation an extrapolation of
Poincare group, and in fact it is the very Weyl group.

A newly similar consideration of the scaling symmetry appears in
Ref. \cite {Hofman11}, in which the authors discuss the scaling
symmetry in 2-dimension system by using the light-cone quantum
field method. And the work \cite {Orland09, Orland11} treat the
scaling transformation based on a first principle form from
Wilsonian method, in which some of the renormalization processes
are repeated. Here we don't follow it in details of
renormalization. We focus more on the application of scaling
feature of renormalization to nonlocal interaction, and also on
what we can infer based on such application. Earlier before there
had been other efforts to associate scaling transformation to
quantum field theory. None of them is satisfactory since, no
perfect quantum system is found so that the corresponding
Lagrangian is scaling invariant unless, the mass of involved
particles are null and, the trace of energy-momentum of the system
becomes zero ~\cite {Jose88,Yu13,Gross70,Kast08,Dirac35,Lus75}. I
think a reason is that they didn't consider the spatial
representation and spinor representation simultaneously. For the
same reason in what follows we derive a conservation law different
from those in literature.

While discussing conservation law, in Lagrangian there are at
least two other additional terms to be involved, namely the
kinetic term $\bar \psi \gamma ^\mu p_\mu \psi $ and mass term
$m\bar \psi (y)\psi (y)$. As for the kinetic term of an extended
particle in the extreme condition, the momentum become light-cone
like and the kinetic mass tends to zero since,
\begin{equation}
m_{kinetic}^2=(\Gamma _\mu p^\mu )(\Gamma _\nu p^\nu )=\gamma _\mu
(1-\gamma
_5)p^\mu \gamma _\nu (1-\gamma _5)p^\nu =p^2(1+\gamma _5)(1-\gamma _5)=0%
\text{ ,}  \label{Kmass}
\end{equation}
here the $m^2=0$ may just have comparable meaning while its
momentum is very large and its mass can be ignored according to
physics. For consistency we prefer to view the kinetic term as the
form $\bar \psi \Gamma ^\mu p_\mu \psi $ and now we know it keeps
invariant under scaling transformation. The invariance of net mass
term is ensured by the following relation if we prefer not to omit
it,
\begin{equation}
m\,\bar \psi (y)\psi (y)=m\,\bar \psi (y)S^{\prime -1}S^{\prime }\psi
(y)=m\,\bar \psi ^{\prime }(y)\psi ^{\prime }(y)\,\,\text{.}  \label{MassC}
\end{equation}
In summary the Lagrangian without mass term yields
\begin{equation}
\mathscr{L}=\bar \psi \Gamma ^\mu p_\mu \psi -gF(q^2)\,A_\mu (q^2)\,j^\mu
(p,p^{\prime }).  \label{Lag}
\end{equation}
where $\,j^\mu (p,p^{\prime })=\bar \psi (p)\Gamma ^\mu \psi
(p^{\prime })$, as the obtained vertex in the last section.

Here we mainly investigate the conserved angular momentum for Eq.
(\ref{Lag}) under the transformation set $\{\{e^{\frac u2\gamma
^\mu \gamma ^\nu }\},e^{\frac u2\gamma _5}\}$. First let's recall
the customarily conserved quantities (Eq. (\ref{AA}) to Eq.
(\ref{Known}) ) under the usual spatial transformation, i.e. the
translations and rotations. These 4-dimensional spatial
transformations with infinitesimal forms are
\begin{equation}
x_\alpha \rightarrow x_\alpha ^{\prime }=x_\alpha +\delta \,x_\alpha
=x_\alpha +\varepsilon _{\alpha \beta }\,x^\beta +\delta _\alpha \,\,,
\label{AA}
\end{equation}
where $\delta _\alpha $ is an infinitesimal displacement and $\varepsilon
_{\alpha \beta }$ is an infinitesimal antisymmetric tensor for rotation in
4-dimension, $\varepsilon _{\alpha \beta }=-\varepsilon _{\beta \alpha }$.
This transformation guarantees the invariance of $x_\alpha x^\alpha $ while $%
\delta _\alpha =0$. The above spatial transformation corresponds to the
transformation for quantum fields as
\begin{equation}
\psi _r(x)\rightarrow \psi _r^{\prime }(x^{\prime })=\psi _r(x)+\frac
12\varepsilon _{\alpha \beta }S_{rs}^{\alpha \beta }\psi _s(x)\,,
\label{Genera}
\end{equation}
in which the matrices elements $S_{rs}^{\alpha \beta }$ are from the spinor
representation of Lorentz group ~\cite{Mandl10}, and in the second term both
of the repeated indices stand for summations, and $\psi _s(x)$'s are
components in $\psi ^T(x)=(\psi _1(x),\psi _2(x),\psi _3(x),\psi _4(x))$.
Take $\delta _\alpha =0$ and impose additionally the invariance of the
Lagrangian
\begin{equation}
\mathscr{L}(\psi _r(x),\psi _{r,\,\alpha }(x))=\mathscr{L}(\psi _r^{\prime
}(x^{\prime }),\psi _{r,\,\alpha }^{\prime }(x^{\prime }))\,,
\end{equation}
one gets a general conserved current (known as the N$\ddot o$ether current)
\cite{Mandl10} relating to angular momentum,
\begin{equation}
j^\alpha =\frac 12\varepsilon _{\beta \gamma }\wp ^{\alpha \beta \gamma }\,,
\label{Current}
\end{equation}
where
\begin{equation}
\wp ^{\alpha \beta \gamma }=\frac{\partial \mathscr{L}}{\partial \psi
_{r,\,\alpha }}S_{rs}^{\beta \gamma }\psi _s+[x^\beta \Im ^{\alpha \gamma
}-x^\gamma \Im ^{\alpha \beta }]\,,  \label{Tensor1}
\end{equation}
and
\begin{equation}
\Im ^{\alpha \beta }=\frac{\partial \mathscr{L}}{\partial \psi _{r,\,\alpha }%
}\frac{\partial \psi _r(x)}{\partial x_\beta }-\mathscr{L}g^{\alpha \beta
}\,.  \label{Tensor2}
\end{equation}
The current Eq. (\ref{Current}) leads to angular momentum operator in four
dimensions by
\begin{equation}
M^{\alpha \beta }=\int \mathbf{d}^{\mathbf{3}}\mathbf{x}\wp ^{0\alpha \beta
}=\int \mathbf{d}^{\mathbf{3}}\mathbf{x}\{[x^\alpha \Im ^{0\beta }-x^\beta
\Im ^{0\alpha }]+\pi _r(x)S_{rs}^{\alpha \beta }\psi _s(x)\}\,,
\label{Known}
\end{equation}
in which $\pi _r(x)=\frac{\partial \mathscr{L}}{\partial \dot \psi _r(x)}$
is a conjugate field of $\psi _r(x)$.

Since the orbital angular momentum is not affected by scaling
transformation due to $[D,M_{\mu \nu }]=0$, we can only add the
new ingredient $\bar \varepsilon _{\mu \nu }\bar S^{\mu \nu
}=\frac{{1}}2\varepsilon \,\gamma _5$ (from $S^{\prime }=e^{\frac
u2\gamma _5}$) into the total variation of field $\psi $ in Eq.
(\ref{Genera}). Thus the spinor part would vary with the change of
$S_{rs}^{\beta \gamma }$, as $\varepsilon _{\beta \gamma
}S_{rs}^{\beta \gamma }\rightarrow \varepsilon _{\beta \gamma
}S_{rs}^{\beta \gamma }+\bar \varepsilon _{\beta \gamma }\bar
S_{rs}^{\beta \gamma }$. We name the latter part ''scalum''. The
conserved current varies correspondingly
\begin{equation}
\tilde j^\alpha =\frac 12\varepsilon _{\beta \gamma }\wp ^{\alpha \beta
\gamma }\,+\frac 12\bar \varepsilon _{\beta \gamma }\bar \wp ^{\alpha \beta
\gamma }\,,  \label{Current2}
\end{equation}
with
\begin{equation}
\bar \varepsilon _{\beta \gamma }\bar \wp ^{\alpha \beta \gamma }=\frac{%
\partial \mathscr{L}}{\partial \psi _{r,\,\alpha }}\bar \varepsilon _{\mu
\nu }\bar S^{\mu \nu }\psi _s=\frac{\partial \mathscr{L}}{\partial \psi
_{r,\,\alpha }}(\frac{{1}}2\varepsilon \,\gamma _5)_{rs}\psi _s\,.
\end{equation}
Since the part $\bar \varepsilon _{\mu \nu }\bar S^{\mu \nu }=\frac{{1}}%
2\varepsilon \,\gamma _5$ is symmetric, to combine it with the
anti-symmetric part $\varepsilon _{\beta \gamma }S_{rs}^{\beta
\gamma }$ and extract a common factor $\varepsilon _{\beta \gamma
}$, we have to multiply a factor $\frac 16\varepsilon _{\beta
\gamma }\varepsilon ^{\beta \gamma }$ ahead of the $\bar
\varepsilon _{\mu \nu }\bar S^{\mu \nu }$. Then Eq. (\ref
{Current2}) yields
\begin{equation}
\tilde j^\alpha =\frac 12\varepsilon _{\beta \gamma }\wp ^{\alpha \beta
\gamma }\,+\frac 1{24}\varepsilon _{\beta \gamma }\varepsilon ^{\beta \gamma
}\frac{\partial \mathscr{L}}{\partial \psi _{r,\,\alpha }}(\varepsilon
\,\gamma _5)_{rs}\psi _s\,,
\end{equation}
and we should caution that in the second term only the product of
the constants $\varepsilon _{\beta \gamma }$ and $\varepsilon $ is
equivalent to infinitesimal constant $\varepsilon _{\beta \gamma
}$ in the first term, in despite of that they have the same
denotations. The left constant $\varepsilon ^{\beta \gamma }$ is a
normal antisymmetric constant satisfying
\begin{equation}
\varepsilon ^{\beta \gamma }=\{
\begin{array}{c}
1\;\text{if }\beta \text{, }\gamma \text{ are different} \\
0\;\text{if }\beta \text{, }\gamma \text{ are the same}
\end{array}
\,,
\end{equation}
Thus apart from the infinitesimal parameter $\varepsilon _{\beta \gamma }$,
the remaining tensor similar to Eq. (\ref{Tensor1}) becomes
\begin{equation}
\tilde \wp ^{\alpha \beta \gamma }=\frac{\partial \mathscr{L}}{\partial \psi
_{r,\,\alpha }}(S_{rs}^{\beta \gamma }+\frac 1{12}w\,\varepsilon ^{\beta
\gamma }(\gamma _5)_{rs})\psi _s+[x^\beta \Im ^{\alpha \gamma }-x^\gamma \Im
^{\alpha \beta }]\,,  \label{FinalTensor}
\end{equation}
the constant $w$ is responsible for the quotient between the two constants $%
\varepsilon _{\beta \gamma }$ and $\varepsilon $, which is assumed to be
adjustable.

Now it is evident that the angular momentum Eq. (\ref{Tensor1})
varies correspondingly with the transformation. In this sense we
conclude that the nonlocal interaction entails the new internal
freedom and becomes the particle intrinsic local property.
Meanwhile it brings about the extrapolation of conventional spin
angular-momentum. To put it in other words, the particles with
shape and those point-like seem to follow different conservation
laws. For the extensive particles it is necessary to involve this
correction term $\frac{\partial \mathscr{L}}{\partial \psi
_{r,\,\alpha }}\frac 1{12}w\,\varepsilon ^{\beta \gamma }(\gamma
_5)_{rs}$ in spin part \cite{Explain}. Thus when an extended
particle (like proton) is smashed we shall not evaluate the
polarizations of its initial state and its final state (smashed
shreds of proton) in conventional way, since the initial state
(proton) and the final states (smashed proton) all have their
non-point size and thus the initial polarization might not be the
sum of its final states (smashed proton) [Fig.2]. Regarding this
different conservation law may help us alleviate the spin crisis
appearing in the polarized electron-nucleon scattering experiment
\cite {FWang1,FWang2,Leader,Ma,Ji12,Ji97A,Ji97B}.

It has been a long-standing puzzle that how the nucleon spin originates from
its constituent parts, namely, the angular momentum of quarks and gluons.
The conflict arose from the estimation of the total spin of proton based on
the experimental value of the antisymmetric structure function $g_1$ ~\cite
{Koda79A,Koda79B,Jaff95}. The total spin $\Sigma $ (defined to be $\Sigma
=\Delta u+\Delta d+\Delta s$, $\Delta u$ is the fraction of $u$ quark in
proton's spin, and the same sense to $\Delta d$ and $\Delta s$) of proton
relates to the structure function $g_1$ by generalizing Bjorken's sum rule ~%
\cite{Bass05}, namely,
\begin{equation}
\int_0^1g_1^p(x,Q^2)=\frac 1{12}[g_A^{(3)}+\frac 13g_A^{(8)}]+\frac 19\Sigma
\,,
\end{equation}
where $g_A^{(3)}=\Delta u-\Delta d$, $g_A^{(8)}=\Delta u+\Delta d-2\Delta s$
are separately the iso-vector, $SU(3)$ octet. $g_A^{(3)}$, $g_A^{(8)}$have
been very well determined respectively from neutron $\beta $-decay and
semi-leptonic hyperon decay ~\cite{Jaff90}. After involving the radiative
correction in perturbative QCD, the above relation can be precisely
interpreted as~\cite{Myhr10}
\begin{equation}
\int_0^1g_1^p(x,Q^2)=\frac{c_1(Q^2)}{12}[g_A^{(3)}+\frac 13g_A^{(8)}]+\frac{%
c_2(Q^2)}9\Sigma \,,  \label{FactorG}
\end{equation}
where the coefficients $c_1(Q^2)$ and $c_2(Q^2)$ come from QCD
perturbative corrections. With above knowledge, hitherto it has
been well known that the value of $\Sigma $ only amounts to $1/3$
of total proton's spin.

People once conceived gluons' spin may contribute much to proton's
spin, but recent experimental analysis~\cite{ExP of Gluon}
supports merely small fraction of gluon's contribution. Another
recent flurry has been the focus on the decomposition of angular
momentum of quark and gluon into spin part and orbital angular
momentum part based on the gauge-invariant QCD dynamics
~\cite{Jaff90,Ji97B,Chen08,Chen09,Waka10,Waka11,Hatta11, Lorc13}.
But the ways to treat angular momentum of bounded quarks are so
controversial that hitherto there has been no widely accepted
scheme. In a very recent paper, Ji et al ~\cite{Ji12} refined the
sum rule using generalized parton
distribution (GPD) method, which may improve further the evaluation of $%
\Sigma $.

Now let's focus on the coefficients of $c_1(Q^2)$ and $c_2(Q^2)$ in Eq. (\ref
{FactorG}), whose accurate values are based on the perturbative calculation
in QCD. For instance, the coefficients $c_1(Q^2)$ reads ~\cite{Bass05}
\begin{equation}
c_1(Q^2)=1-(\frac{\alpha _s}\pi )-3.58333(\frac{\alpha _s}\pi )^2+\cdots \,,
\label{Coeff}
\end{equation}
in which $\alpha _s(Q^2)$ may have a running value with respect to
$Q^2$, roughly around 0.1\symbol{126}0.3. Because the scale
transformation is somehow derived from the renormalization group,
including the running of charges etc., one may be aware of that
the corrections in Eq. (\ref{Coeff}) are to some extent equal to
the scaling transformation. And the corrections from Eq.
(\ref{Coeff}) might also be consistent with effect that
interpreted by Eq. (\ref{FinalTensor}). Though the corrections
have not made the coefficients $c_1(Q^2)$ and $c_2(Q^2)$ deviate
so much from $1$, we know the corrections actually affect the
value of $\Sigma $. When the scale approaches to the
nonperturbative regime and $\alpha _s(Q^2)$ becomes larger, the
expression of Eq. (\ref{Coeff}) however, may lose its validation.
Whereas we note our scaling transformation happens to be
responsible for the shift between the perturbative and
nonperturbative regimes since the dilation (shrinkage) occurs
accompanying with the loss (injection) of energy. $Nu$
$\rightarrow \pm \infty $ may imply the coupling constant $\alpha
_s\rightarrow \infty $ according to our previous understanding of
the form factor, so the extreme states are really relevant. We
thus speculate that the scaling transformation might be helpful to
transform the spin value from perturbative scale to
nonperturbative scale, or vice versa. In this sense the
transformation method could be a way to find an explanation on the
spin crisis of proton. And further investigation is in progress.
While the SU(3) group steps in, some unexpected effects may occur.

\section{The impact of vertex $F(q^2)A_\mu (k)\bar \psi (p)\gamma ^\mu
(a+b\,\gamma _5)\psi (p)$ on polarized scattering}

In this section we will discuss that after finite steps of scaling
transformations, what is the contribution of the evolving vertex
$A_\mu (k)\bar \psi (p)\gamma ^\mu (a+b\,\gamma _5)\psi (p)$ to
the scattering processes. Analogous to the inelastic e-p
scattering, where the assumed vertex $A\gamma ^\mu +B\,\sigma
^{\mu \nu }p_\nu $ yields some observed structures $W^{\mu \nu }$
in cross-section \cite{Greiner}, here we are concerned about what
the structures the evolving vertex-form $\gamma ^\mu (a+b\,\gamma
_5)$ would lead to. Although we work following the analogy, we
should caution that we focus on elastic scattering, rather than
inelastic scattering. At the end of this section we arrive at the
conclusions that the part $b\,\gamma ^\mu \gamma _5$ contributes
nothing to the normal unpolarized cross-section of elastic
scattering, so effectively it doesn't change the conventional
structure form. However, for the polarized scattering, there
appears exceptional terms additional to the original structure
function.

Firstly, let's carry out the structure function of unpolarized cross-section
by averaging over the initial spins and summing over the final spins \cite
{Greiner}. Without losing generality, let's suppose that it is the very case
for an electron (a point fermion) incident on nucleon (an extended fermion)
[Fig.3]. By conventional steps, one finds that the evolving vertex-form $%
\gamma ^\mu (a+b\,\gamma _5)$ yields the following scattering tensor,
\begin{equation}
W_{\mu \nu }^{(p)}=\frac 12\text{Tr}[\frac{q_2+M}{2M}\gamma _\mu
(a+b\,\gamma _5)\frac{p_2+M}{2M}\gamma _\nu (a+b\,\gamma _5)]\,,
\end{equation}
and apart from the coefficient $\frac 12$, the trace can be separated into
four terms

\begin{equation}
\begin{tabular}{l}
$W_{\mu \nu }^{(p)}=W_{\mu \nu }^{(1)}(a^2)+W_{\mu \nu }^{(2)}(a\,b)+W_{\mu
\nu }^{(3)}(b\,a)+W_{\mu \nu }^{(4)}(b^2)$ \\
$=\frac 12a^2\,\text{Tr}[\frac{q_2+M}{2M}\gamma _\mu \frac{p_2+M}{2M}\gamma
_\nu ]+\frac 12a\,b\,\text{Tr}[\frac{q_2+M}{2M}\gamma _\mu \frac{p_2+M}{2M}%
\gamma _\nu \gamma _5]$ \\
$+\frac 12b\,a\,\text{Tr}[\frac{q_2+M}{2M}\gamma _\mu \gamma _5\frac{p_2+M}{%
2M}\gamma _\nu ]+\frac 12b^2\,\text{Tr}[\frac{q_2+M}{2M}\gamma _\mu \gamma _5%
\frac{p_2+M}{2M}\gamma _\nu \gamma _5]\text{ ,}$%
\end{tabular}
\end{equation}
and the result of the first term is well-known, it is
\begin{equation}
W_{\mu \nu }^{(1)}(a^2)=\frac{a^2}{2\,M^2}[q_{2\mu }p_{2\nu }+q_{2\nu
}p_{2\mu }-(q_2\cdot p_2-M^2)g_{\mu \nu }]\,,  \label{ScaTenA}
\end{equation}
the second term is
\begin{equation}
W_{\mu \nu }^{(2)}(a\,b)=\frac{a\,b}{\,M^2}i\varepsilon _{\mu \nu \rho
\sigma }q_2^\rho p_2^\sigma \,,  \label{ScaTenB}
\end{equation}
and the third term results in the same
\begin{equation}
W_{\mu \nu }^{(3)}(b\,a)=\frac{a\,b}{\,M^2}i\varepsilon _{\mu \nu \rho
\sigma }q_2^\rho p_2^\sigma \,.  \label{ScaTenC}
\end{equation}
The last term has the same form as the first term, $W_{\mu \nu }^{(1)}(a^2)$%
, apart from the coefficient
\begin{equation}
W_{\mu \nu }^{(4)}(b^2)=\frac{b^2}{2\,M^2}[q_{2\mu }\,p_{2\nu }+q_{2\nu
}\,p_{2\mu }-(q_2\cdot p_2-M^2)\,g_{\mu \nu }]\,.  \label{ScaTenD}
\end{equation}
We note the new structure functions are from Eqs. (\ref{ScaTenB},
\ref
{ScaTenC}), whose contribution however, would vanish since the tensor $%
\varepsilon _{\mu \nu \rho \sigma }$ are antisymmetric, while the tensor of
lepton part $L^{\mu \nu }$ coupled to them is symmetric with respect to
indices $\mu, \nu $.

Secondly, let's consider the polarized cross-section. Now we do
not fix the initial or the final spin states and leave the spin
operator in the potential. With the same marks as in Fig.3 and
without the propagator, we write the polarized amplitude
(potential) by using the Dirac spinors as follows
\begin{equation}
\begin{tabular}{l}
$M_{AV}(\vec p,\vec q,\vec k)=\bar \psi _1\gamma _{1\mu }(a+b\,\gamma
_5)\psi _1\,\bar \psi _2\gamma _2^\mu (a+b\,\gamma _5)\psi _2$ \\
$=a^2\,\bar \psi _1\gamma _{1\mu }\psi _1\,\bar \psi _2\gamma _2^\mu \psi
_2+a\,b\,\bar \psi _1\gamma _{1\mu }\psi _1\,\bar \psi _2\gamma _2^\mu
\,\gamma _5\psi _2$ \\
$+b\,a\,\bar \psi _1\gamma _{1\mu }\,\gamma _5\psi _1\,\bar \psi _2\gamma
_2^\mu \psi _2+b^2\,\bar \psi _1\gamma _{1\mu }\,\gamma _5\psi _1\,\bar \psi
_2\gamma _2^\mu \,\gamma _5\psi _2\text{ ,}$%
\end{tabular}
\end{equation}
Only the terms with coefficients $ab$ and $ba\,$(henceforth we
denote the two terms as $\{ab\}$ and $\{ba\}$) are new, and the
results for the other
two terms $\{aa\}$ and $\{bb\}$ can be found in Ref. \cite{Su92}. The term $%
\{ab\}$ can be tidied up into
\begin{equation}
\begin{tabular}{l}
$M_{\{a\,b\}}=\bar U(\vec p_1)\gamma _{1\mu }U(\vec q_1)\bar U(\vec
p_2)\gamma _2^\mu \,\gamma _5U(\vec q_2)$ \\
$=\bar U(\vec p_1)\gamma _{10}U(\vec q_1)\bar U(\vec p_2)\gamma _2^0\,\gamma
_5U(\vec q_2)-\bar U(\vec p_1)\vec \gamma _1U(\vec q_1)\cdot \bar U(\vec
p_2)\vec \gamma _2\,\gamma _5U(\vec q_2)$ ,
\end{tabular}
\end{equation}
where the indices 1, 2 represent respectively the first and the
second
particles. Substitute the concrete form of Dirac spinor $U(\vec p)=\sqrt{%
\frac{E+m}{2\,E}}\left(
\begin{tabular}{l}
$1$ \\
$\frac{\vec \sigma \cdot \vec p}{E+m}$
\end{tabular}
\right) $ and $\bar U(\vec p)=U^{\dagger }(\vec p)\gamma _0$ (where $E=\sqrt{%
\vec p^2+m^2}$) into the above equation, and after lengthy calculation, it
yields
\begin{equation}
\begin{tabular}{l}
$M_{\{a\,b\}}=A\{\frac{\vec \sigma _2\cdot \vec p_2}{E(\vec p_2)+m_2}+\frac{%
\vec \sigma _2\cdot \vec q_2}{E(\vec q_2)+m_2}+\frac{\vec \sigma _1\cdot
\vec p_1}{E(\vec p_1)+m_1}\frac{\vec \sigma _1\cdot \vec q_1}{E(\vec q_1)+m_1%
}\frac{\vec \sigma _2\cdot \vec p_2}{E(\vec p_2)+m_2}$ \\
$+\frac{\vec \sigma _1\cdot \vec p_1}{E(\vec p_1)+m_1}\frac{\vec \sigma
_1\cdot \vec q_1}{E(\vec q_1)+m_1}\frac{\vec \sigma _2\cdot \vec q_2}{E(\vec
q_2)+m_2}-[\frac{\vec \sigma _1\cdot \vec p_1}{E(\vec p_1)+m_1}{\underline{%
\vec \sigma }_1}\frac{\vec \sigma _2\cdot \vec p_2}{E(\vec p_2)+m_2}{%
\underline{\vec \sigma }_2}\frac{\vec \sigma _2\cdot \vec q_2}{E(\vec
q_2)+m_2}$ \\
$+\frac{\vec \sigma _1\cdot \vec p_1}{E(\vec p_1)+m_1}\vec \sigma _1\cdot
\vec \sigma _2+{\underline{\vec \sigma }_1}\frac{\vec \sigma _1\cdot \vec q_1%
}{E(\vec q_1)+m_1}{\underline{\vec \sigma }_2}+{\underline{\vec \sigma }_1}%
\frac{\vec \sigma _1\cdot \vec q_1}{E(\vec q_1)+m_1}\frac{\vec \sigma
_2\cdot \vec p_2}{E(\vec p_2)+m_2}{\underline{\vec \sigma }_2}\frac{\vec
\sigma _2\cdot \vec q_2}{E(\vec q_2)+m_2}]\}\text{ ,}$%
\end{tabular}
\end{equation}
here we use underlines to denote the inner product and
\begin{equation}
A=\sqrt{\frac{E(\vec p_1)+m_1}{2\,E(\vec p_1)}\frac{E(\vec p_2)+m_2}{%
2\,E(\vec p_2)}\frac{E(\vec q_1)+m_1}{2\,E(\vec q_1)}\frac{E(\vec q_2)+m_2}{%
2\,E(\vec q_2)}}\approx 1-\frac 1{8m_1^2}(\vec p_1^2+\vec q_1^2)-\frac
1{8m_2^2}(\vec p_2^2+\vec q_2^2)\,.
\end{equation}
The second step follows while using the approximation $E(\vec p_1)\approx
m_1+\frac{\vec p_1^2}{2m_1}$ and to the order of $\frac{\vec p^2}{m^2}$.
With this approximation, one further gets
\begin{equation}
\begin{tabular}{l}
$M_{\{a\,b\}}=A\{\frac 1{2\,m_2}\vec \sigma _2\cdot (\vec p_2+\vec
q_2)-\frac 1{2\,m_1}[\vec \sigma _1\cdot \vec p_1\vec \sigma _1\cdot \vec
\sigma _2-{\underline{\vec \sigma }_1}\vec \sigma _1\cdot \vec q_1{%
\underline{\vec \sigma }_2}]\}$ \\
$=A\{\frac 1{2\,m_2}\vec \sigma _2\cdot (\vec p_2+\vec q_2)-\frac
1{2\,m_1}[(\vec p_1-\vec q_1)\cdot \vec \sigma _2-i(\vec \sigma _1\times
\vec \sigma _2)\cdot (\vec p_1+\vec q_1)]\}$ ,
\end{tabular}
\label{LastA}
\end{equation}
in the second step of the above equation we have used the following
relations
\begin{equation}
\sigma _i\sigma _j=\delta _{ij}+i\varepsilon _{ijk}\sigma _k\,,
\label{LLastA}
\end{equation}
and
\begin{equation}
(\vec \sigma \cdot \vec A)(\vec \sigma \cdot \vec B)=A_iB_j\sigma _i\sigma
_j=\vec A\cdot \vec B+i(\vec A\times \vec B)\cdot \vec \sigma \,.
\label{LLastB}
\end{equation}
Likewise, we obtain $M_{\{ba\}}$ as follows
\begin{equation}
M_{\{ba\}}\simeq A\{\frac 1{2\,m_1}\vec \sigma _1\cdot (\vec p_1+\vec
q_1)-\frac 1{2\,m_2}[(\vec p_2+\vec q_2)\cdot \vec \sigma _1+i(\vec \sigma
_1\times \vec \sigma _2)\cdot (\vec p_2-\vec q_2)]\}\,.  \label{LastB}
\end{equation}

We note that the terms $M_{\{a\,b\}}$ and $M_{\{ba\}}$ do not
appear in those cross-sections derived from any single of five
Lorentz-invariant currents $\bar \psi (p)\psi (p)$, $\bar \psi
(p)\gamma ^\mu \psi (p)$, $\bar
\psi (p)\,\gamma _5\psi (p)$, $\bar \psi (p)\gamma _5\gamma ^\mu \psi (p)$, $%
\bar \psi (p)\gamma ^\mu \gamma ^\nu \psi (p)$, unless some of
them are mixed. Thus we realize that $M_{\{a\,b\}}$ and
$M_{\{ba\}}$ can actually occur if the current is the weak current
$\bar \psi (p)\gamma ^\mu (1-\,\gamma _5)\psi (p)$, for instance
in the scattering of neutrino incident on electron. However, since
the intermediate Z boson is very heavy and thus the scattering
involving weak interaction only appears in very high energy, we
may avoid the case by testing effects of nonlocality in somewhat
lower energy regime. Moreover, mostly the vertex we meet in
nonlocal current must be $\bar \psi (p)\gamma ^\mu (a+\,b\gamma
_5)\psi (p)$ instead of pure extremes $\bar \psi (p)\gamma ^\mu
(1\pm \,\gamma _5)\psi (p)$, which leave coefficients $a$, $b$ to
adjust. Unexpectedly, if the nonlocal extreme vertices were mixed
or entangled with weak interaction and were evidenced by
experiments, then it must be a most intriguing topic deserving
further investigation.

We would like to present a simple gedanken experiment to test
nonlocal effect due to the handedness terms, which are all
proportional to helicity $\vec \sigma \cdot \vec p$ in one manner
or another, as shown in Eqs. (\ref {LastA}, \ref{LastB}). For the
feasibility of the experiment, we turn from hadron dynamics to
molecular scale to test the prediction of Eqs. (\ref{LastA},
\ref{LastB}). Imagine that an electron scattered away from a
simple atom like hydrogen, which stays in its ground state, so
that no orbital angular momentum is involved in the scattering
processes. Meanwhile we should control the energy of incident
electron to be low enough so that other orbital states of
hydrogen-atom are not involved. Maybe the energy should $E\leq
1eV$ or several $eV$s (with wavelength less than $2\stackrel{
\circ }{A}$), which is largely lower than its first threshold of
transitions. The electron's energy have to be controlled precisely
to limit its wavelength less than hydrogen-atom diameter and
meanwhile not so short as to cause transition of hydrogen-atom. If
actually the energy of the electron is not easy to control, we may
directly use the scattering between ground-state hydrogen atoms
instead of the scattering between electron and hydrogen-atom. In
such scenario the total angular momentum of hydrogen atom is its
spin, and the nucleon magnetic moment is omitted for its small
fraction in the total (about 1 in 2000). And the spreading
electron cloud of hydrogen atom meets the case of our nonlocal
description. Such hydrogen atom could be good testing ground for
nonlocal predictions.

The proposed experiment is to use polarized electrons (or hydrogen atoms)
colliding on polarized hydrogen-atom (polarized by magnetic field).
Different from the previous calculations on polarized electron-electron
(e-e) scattering \cite{Wann07}and e-H scattering~\cite
{Blum75,Wyng86,Kels79,Edmu83,Phil62,Ohm58}, the main results there are shown
in Eqs. (\ref{LastA}, \ref{LastB}), which is characterized by terms like $%
i(\vec \sigma _1\times \vec \sigma _2)\cdot (\vec p_2-\vec q_2)$. Such term
differs from the normal handedness term $\vec \sigma \cdot \vec p$ in that
it permits the existence of two perpendicular spins $\vec \sigma _1$ and $%
\vec \sigma _2$. Whereas the previous test-experiments on spin asymmetry
\cite{Wyng86} mainly focused on the parallel or anti-parallel difference, as
\begin{equation}
A_{1s,1s}(\theta )=\frac{\sigma _{\uparrow \downarrow }-\sigma _{\uparrow
\uparrow }}{\sigma _{\uparrow \downarrow }+\sigma _{\uparrow \uparrow }}%
\text{ ,}  \tag{43a}
\end{equation}
or
\begin{equation}
A_{1s,1s}(\theta )=\frac{\sigma _{\Rightarrow }-\sigma _{\Leftarrow }}{%
\sigma _{\Rightarrow }+\sigma _{\Leftarrow }}\text{ ,}  \tag{43b}
\end{equation}
where $\sigma _{\uparrow \downarrow }$, $\sigma _{\uparrow \uparrow }$, or $%
\sigma _{\Rightarrow }$, $\sigma _{\Leftarrow }$ mean the parallel or
anti-parallel in common sense. However, in our case, we suggest a new
asymmetric parameter
\begin{equation}
A_{1s,1s}(\theta )=\frac{\sigma _{\rightarrow \uparrow }-\sigma _{\uparrow
\rightarrow }}{\sigma _{\rightarrow \uparrow }+\sigma _{\uparrow \rightarrow
}}\text{ ,}  \tag{43c}
\end{equation}
which has never been investigated in previous theoretical study \cite
{Blum75,Wyng86,Kels79,Edmu83,Phil62,Ohm58} or experiments \cite
{Crowe90,Flet82,Will75,Gay82}. But this term may contribute even smaller
fractions to total cross-section, since the sum of all spin-dependent terms
have attributed just minor fraction in total cross-section. The calculation
details can follow the paper \cite{Wann07}, concerning additionally the
present interaction. If any experiment gets a nontrivial parameter $\frac{%
\sigma _{\rightarrow \uparrow }-\sigma _{\uparrow \rightarrow }}{\sigma
_{\rightarrow \uparrow }+\sigma _{\uparrow \rightarrow }}$ then it proves
our predictions. And maybe each of the three parameters ($\frac{\sigma
_{\uparrow \downarrow }-\sigma _{\uparrow \uparrow }}{\sigma _{\uparrow
\downarrow }+\sigma _{\uparrow \uparrow }}$ , $\frac{\sigma _{\Rightarrow
}-\sigma _{\Leftarrow }}{\sigma _{\Rightarrow }+\sigma _{\Leftarrow }}$, and
$\frac{\sigma _{\rightarrow \uparrow }-\sigma _{\uparrow \rightarrow }}{
\sigma _{\rightarrow \uparrow }+\sigma _{\uparrow \rightarrow }}$) would
deviate from former evaluations in literature if involving all of the
handedness terms $\vec \sigma \cdot \vec p$ in Eqs. (\ref{LastA}, \ref{LastB}%
).

\section{Conclusions and Discussions}

In this paper we have discussed elaborately the role of scaling
transformation in nonlocal interaction. This transformation
pertains to describing the relationship of different
energy/space-time scales in scattering between (fermion) hadrons.
The scaling transformation is recognized/constructed based on the
conclusions of RGM and the popular expressions of conformal group.
The most significant feature of this paper is to combine its
spinor representation $\gamma _5$ and coordinate representation
$i\,x_\mu
\partial ^\mu $ together. To this end, we surmise there is a local
vertex $\Gamma ^\mu $ transforming as $S^{^{\prime }-1}\Gamma ^\mu
S^{\prime }=\Lambda _{\;\nu }^{\prime \mu }\Gamma ^\nu $, in which
$S^{\prime }=e^{\frac u2\gamma _5}$, resembling Lorentz
transformation acting on vector vertex, $ S^{-1}\gamma ^\mu
S=\Lambda _{\;\nu }^\mu \gamma ^\nu $, where $S$ corresponds to
spinor representation of Lorentz transformation. In this way we
obtain the scaling invariant vertices $\Gamma ^\mu =\gamma ^\mu
(1\pm \gamma _5)$, which means the invariance of interaction
vertex $A_\mu (x)\bar \psi (x)\Gamma ^\mu \psi (x)$ while
performing the scaling transformation.

~~\\

Based on the knowledge that the transformation $S^{\prime }$ is applied
repeatedly to vector vertex $e^{-\frac u2\gamma _5}\gamma ^\mu e^{\frac
u2\gamma _5}\simeq \gamma ^\mu +u\,\gamma ^\mu \gamma _5=(1-u\,)\gamma ^\mu
+u\,\gamma ^\mu (1+\gamma _5)=(1+u\,)\gamma ^\mu -u\,\gamma ^\mu (1-\gamma
_5)$, one finds the varying coefficients ahead of $\gamma ^\mu $%
, which matches the running coupling constant occurring in RGM. As
for vertices $\gamma _\mu (1\pm \gamma _5)$, here they are viewed
as extremes of normal vector vertex-form $\gamma ^\mu $ after
infinite steps of scaling transformation since, $Nu$ $\rightarrow
\pm \infty $, $(e^{-\frac u2\gamma _5})^N\gamma ^\mu (e^{\frac
u2\gamma _5})^N=\gamma ^\mu \cosh Nu+\gamma ^\mu
\gamma _5\sinh Nu\rightarrow \gamma _\mu (1\pm \gamma _5)$. We also call $%
\gamma _\mu (1\pm \gamma _5)$ the vertices at extreme condition,
which might be the system of very high energy or at very low
temperature. We further discuss the conservation law for these
extreme vertices, for which an extra intrinsic-degree named scalum
is introduced into the total angular momentum. Based on the
experience from renormalization, the parameter $\mu $ is somewhat
equivalent to such a degree of freedom. It is natural for us to
associate the results of conservation law with the spin crisis of
nucleons, responding to the appearance of the scalum.

~~\\

The extreme states as well as the extreme vertex may not exist in
nature. However, by the inquiring and inferring process we
recognize that the conformal group exists more like for running
properties rather than for invariance of quantum fields. For an
extended particle involved in a scattering at certain energy, we
have to make corresponding scaling transformations to interpret
locally its interaction vertex. Assume a nonlocal interaction
interpreted initially/unphysically by exchanging vector bosons,
then a general interaction-vertex $\,\gamma ^\mu (a+b\,\gamma _5)$
exists, with which we use local vertex-form to interpret the
nonlocal interaction. The general interaction-vertex has effects
on polarized scattering rather than unpolarized scattering.
Accordingly we propose a gedanken experiment to test our
predictions on nonlocal interaction. The experiment is based on
the scattering between a charged point-particle and the ground
state of a hydrogen. That is recognized as a good method to test
nonlocal interaction-vertex, since the cloud of ground-state
electron distributes around the nucleon so that it forms a
nonlocal region, meanwhile all of its angular momentum is the spin
of the electron.

~~\\

Although the dynamics used in this paper mostly stems from the
perturbative dynamics, it opens a door for our understanding to
nonperturbative dynamics. There have been continuous efforts to
study nonperturbative interaction ever since the birth of
renormalization~\cite{Ber02, Mas08}. To apply somehow the scale
parameter of renormalization to intermediate-strong-interaction
was the primary goal of this paper. Furthermore an even stronger
motivation is to develop an analytic non-perturbation method to
understand such intermediate-strong-interaction. The motivation
has driven us to apply transformation instead of solely the scale
parameter to nonperturbative interactions. Some other nonlocal
theories have made efforts to link the nonlocal interaction with
renormalization, for instance in Ref. \cite{Moffat89, Moffat90,
Efimov67, Efimov72}. But none of them used transformation method,
which was laid there years before \cite{Green63, Taka71}. In our
results, the appearance of $\gamma _5$ in both the scaling
transformation and the nonlocal vertex-form gives us the
confidence that we might have unveiled a truth of nonperturbative
dynamics. Since when a current quark gains its mass
non-perturbatively to become a constituent quark, the $\gamma _5$
as well as the chiral-symmetry breaking would occur. Next if
possible we aim to construct a general description of
nonperturbative systems based on their nonlocal properties.

~\\

\begin{acknowledgments}
I am grateful to the hospitality of Prof. Y. B. Dong and Prof. P.
Wang when I visited High Energy Institute of Chinese Academy, as
well as the fruitful discussions with them. Also thanks for the
encouragements from Prof. W. Q. Wang and Prof. W. Han. The Project
Sponsored by the Scientific Research Foundation for the Returned
Overseas Chinese Scholars, State Education Ministry and
Fundamental Research Funds for the Central Universities.
\end{acknowledgments}

~~\\

~~\\

\textbf{Captions}

~~\\

Fig1. The physics picture of performing finite steps of scaling
transformation.
~~\\

Fig2. Schematic diagram for the colliding process between a point
particle and an extended particle.

~~\\

Fig3. The Feynman graph for calculating the scattering
cross-section for point particles, with vertex-form $A_\mu \gamma
^\mu (a+b\gamma ^5)$.


\begin{thebibliography}{999}
\bibitem{Explain}  Scalum turns out to be a new inner degree of freedom.
Yukawa and Yennie~\cite{Yukawa50, Yennie50} once contemplated that
there should appear new ''inner motion'' degree of freedom in a
nonlocal system, later it proved to be trivial due to their
wrongly interpreting manner.
\end{thebibliography}
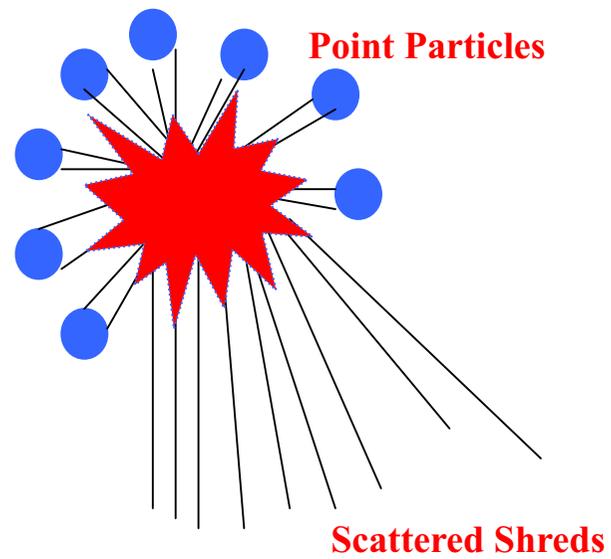
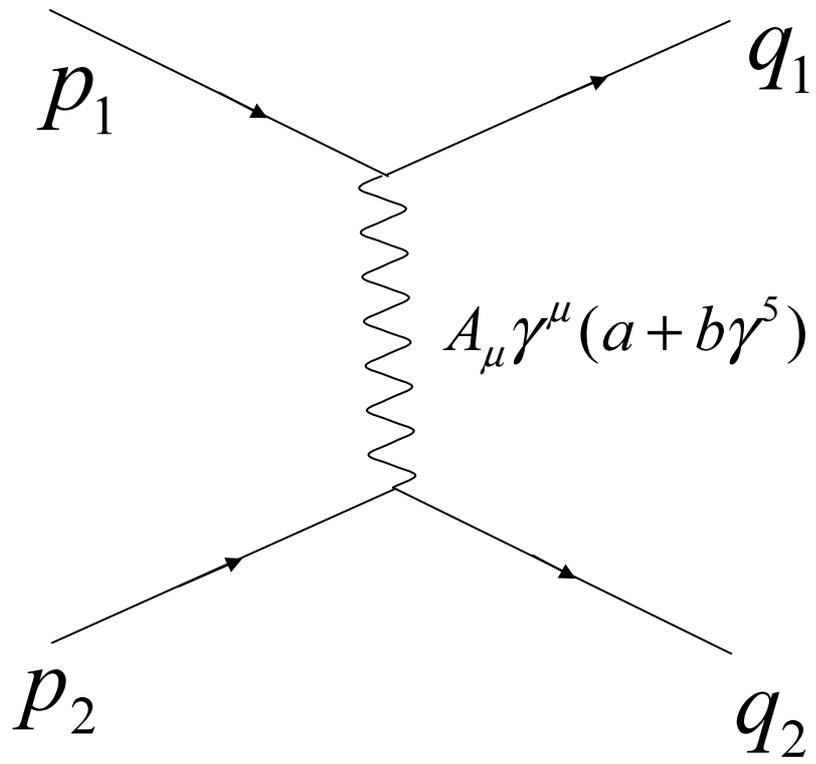
\end{document}